# Coherently excited nonlocal quantum features using polarization-frequency correlation via a quantum eraser

Byoung S. Ham
School of Electrical Engineering and Computer Science, Gwangju Institute of Science and Technology
123 Chumcangwagi-ro, Buk-gu, Gwangju 61005, S. Korea
(Submitted on June 09, 2022, bham@gist.ac.kr)

**Abstract:**
Indistinguishability is an essential concept to understanding "mysterious" quantum features in the view point of the wave-particle duality of quantum mechanics. The fundamental physics of the indistinguishability lies in quantum superposition of a single photon via orthogonal bases in a Hilbert space. Here, a pure coherence approach is applied to the nonlocal correlation using coherent photons manipulated for polarization-frequency correlation. For this, both wave mixing and heterodyne detection techniques are applied for the delayed-choice experiments of a quantum eraser using coherent photons to selectively choose entangled photon pair-like inseparable tensor, otherwise resulting in a typical classical bound with 50% visibility in nonlocality. Thus, the mysterious quantum feature of nonlocal correlation is now coherently understood and may open the door to macroscopic quantum information processing.

**Introduction**
In quantum mechanics, measurements affect the original quantum state [1,2]. Measurement-based quantum mystery has been studied for various delayed-choice experiments over the last several decades [3-12]. Not only a single photon-based quantum superposition [4-7], but also two photon-based nonlocal correlation [10-13] has been intensively studied for the mysterious quantum feature violating causality. Potential quantum loopholes in such a quantum system have also been closed for detection [13,16], locality [14], sampling [13,15,16], and free will [17] parameters. The nonlocal correlation between space-like separated local measurements results in nonlocal realism of quantum mechanics [18]. Classical realism is for an independent reality not influenced by the action of measurements as claimed by Einstein and his colleagues [19]. Nonlocal realism defies the classical reality via quantum features of "spooky action at a distance" [20]. This nonlocal correlation violating local realism has been understood as a unique feature of quantum mechanics that cannot be explained or achieved by any classical means. To satisfy the nonlocal correlation, it is commonly accepted that target photons must be nonclassical [13-23].

  Here in this paper, a contradictory idea of nonlocal correlation is presented by classical means of coherence optics. For this, a coherently excited Franson-type nonlocal quantum feature is studied for the inseparable polarization basis-products via a quantum eraser of the delayed-choice experiments. For the inseparable basis products, coherent photons are manipulated to be frequency-polarization correlated by acousto-optic modulators (AOMs) for wave mixing and its measurements by a heterodyne detection technique. The heterodyne detection should satisfy coincidence detection between two remote local detectors in a microscopic regime, but can also be applied to a macroscopic realm of coherent fields such as continuous waves of a laser. For the heterodyne detection, the temporal resolving time of a photon detector must be shorter than the inverse of the difference frequency between two input photons.

  Recently, the quantum feature of Franson-type nonlocal correlation [24-30] has been newly interpreted with the particle nature of photons using coincidence detection, where the temporal measurement choices of nonlocal correlation violate local realism of the hidden variable theory [31]. In this interpretation, inseparable basis products of nonlocal photon pairs can be accomplished via post-selection of measurement events in a usual coincidence detection manner, resulting in the second-order quantum superposition between the same basis products. On the other hand, the Wheeler's delayed-choice experiments [3-12] have also been newly interpreted for the post-measurement control of polarization basis using a pure coherence approach, resulting in causality violation even to a macroscopic regime [32]. Unlike the Franson correlation [24-30], the delayed choice is for the first-order quantum superposition of a single photon in an interferometric system. The coherent photon-excited inseparable basis-product can be excited by delayed-choice experiments of a quantum eraser [32,33], where the coherent photons are manipulated to be frequency-polarization correlated using AOMs and a polarization beam splitter (PBS). Unlike the particle nature-based inseparable basis-product superposition via coincidence detection [24-30], the present coherence approach of the wave nature also generates the same quantum feature via heterodyne detection, resulting in selective choices of difference frequency pairs. To understand this coherently excited nonlocal correlation, a coherence feature between paired photons via heterodyne detection is analyzed for both local and coincidence detections via post-control of the polarization-basis measurements, satisfying the quantum eraser.

**Results**

*Coherently excited polarization-frequency correlation*

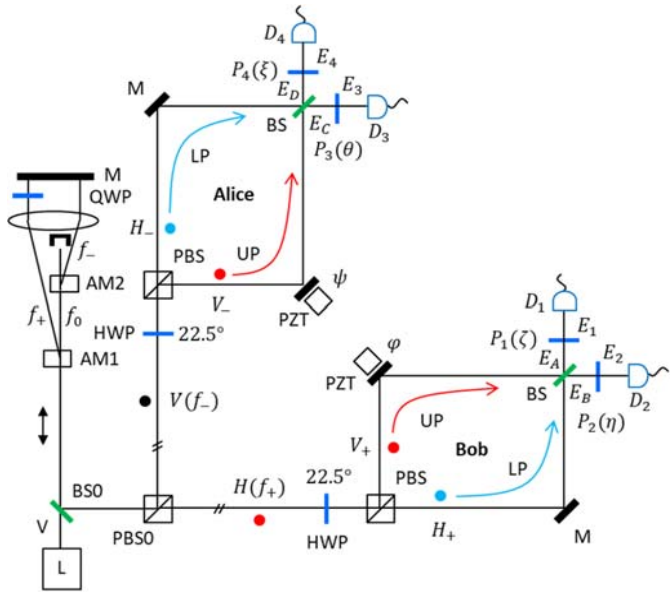

**Fig. 1. Schematic of coherence version of Franson-type nonlocal correlation.** AM: acousto-optic modulator, BS: beam splitter, D: single photon detector, H (V): horizontal (vertical) polarization, HWP: half-wave plate, L: laser, M: mirror, PBS: polarizing BS, P: polarizer, PZT: piezo-electric transducer, QWP: quarter-wave plate. The black (red) dots indicates a paired single photon, whose polarization and frequency are vertical (horizontal) and $f_-$ ($f_+$), respectively.

Figure 1 shows schematic of the coherently excited Franson-type nonlocal correlation using an attenuated laser via a quantum eraser of the delayed-choice experiments. For this, a pair of double-pass AOMs is used for difference frequency generations correlated with polarization bases, satisfying both $f_+ - H$ and $f_- - V$ relationship via a PBS. For the orthogonal polarization bases, a quarter-wave plate (QWP) is placed for only one AOM. To maximize the generation efficiency, the beam splitter BS0 may be set for a high extinction ratio of reflection to transmission. The generated polarization-frequency pair is split into two different paths toward Alice and Bob. To satisfy nonlocality, the path distance between PBS and each MZI is set to be space like, where the light cone should violate causality. In each path, the paired photons random in polarization bases via a 22.5°-rotated half-wave plate (HWP), resulting in a random polarization basis configuration for local measurements. Due to orthogonal polarization bases, each MZI results in no fringe in the output measurements due to Fresnel-Arago law [34], unless a polarizer P is added. The function of P is to induce a quantum eraser [32]. For a diagonally or anti-diagonally rotated P, perfect random polarization bases are provided in each MZI. Thus, the photon characteristic in each MZI satisfies indistinguishability in the view point of the particle nature of quantum mechanics. Due to independence between polarizers and MZIs, any violation of the causality witnesses the "delayed choice," resulting in the fundamental quantum feature [3-13]. As discussed in Ref. [32], the P-based quantum eraser is the origin of Bell inequality violations [10,33].

Regarding the light source in Fig. 1, Poisson-distributed single photons are taken from an attenuated continuous wave (cw) laser. In general, however, there is no difference between a single photon and continuous wave (cw) light for the MZI fringes due to self-interference [35,36]. Nevertheless, the mean photon number is set extremely low to be $\langle n \rangle \sim 0.04$, satisfying the incoherence condition between consecutive single photons. For this, the consecutive photon-to-photon distance in each MZI is set to be much longer than the laser's coherence length. Unlike SPDC-based entangled photons, whose detuning is symmetric across the center frequency [37], coherent photons have no such detuning mechanism. The symmetric frequency detuning in SPDC-generated photon pairs is the major requirement for the Franson-type nonlocal correlation [31].

*Inseparable second-order intensity correlation using coincident heterodyne detection*

According to coherence optics based on single photons in Fig. 1 (see the red and black dots), each photon's amplitude for Alice and Bob is denoted by $E_0$. Due the thumb rule of quantum mechanics that a photon never interferes with others by Dirac [36], a specific number of photons do not matter for the delayed-choice experiments [32]. However, to satisfy the coincidence detection for the nonlocal correlation, the input photons must be doubly bunched to split into both paths by PB0. In this case, 50 % of non-split photons does also contribute to the coincidence detection, but conditionally eliminated by the heterodyne detection method. For Fig. 1, the output photons from both noninterfereing MZIs can be represented as:

$$\begin{bmatrix} E_A \\ E_B \end{bmatrix} = \frac{E_0}{\sqrt{2}} e^{i\eta} \begin{bmatrix} H_+ - V_+ e^{i\varphi} \\ i(H_+ + V_+ e^{i\varphi}) \end{bmatrix}, \quad (1)$$

$$\begin{bmatrix} E_C \\ E_D \end{bmatrix} = \frac{E_0}{\sqrt{2}} \begin{bmatrix} H_- - V_- e^{i\varphi} \\ i(H_- + V_- e^{i\varphi}) \end{bmatrix}, \quad (2)$$

where $H_\pm$ and $V_\pm$ denote a unit vector of horizontal and vertical polarizations, which are correlated to the frequencies $f_\pm$, respectively. The global phase $e^{i\eta}$ in equation (1) is due to the path-length difference from PBS0 to both MZIs. Equations (1) and (2) are for the first-order amplitudes of a single photon, resulting in no fringes due to PBSs, satisfying distinguishable photon characteristics of the MZI.

By the action of P-based post-measurements, Eqs. (1) and (2) are rewritten for a quantum eraser as [32]:

$$\begin{bmatrix} E_1 \\ E_2 \end{bmatrix} = \frac{E_0}{\sqrt{2}} e^{i\eta} \begin{bmatrix} H_+ cos\zeta - V_+ sin\zeta e^{i\varphi} \\ i(H_+ cos\eta + V_+ sin\eta e^{i\varphi}) \end{bmatrix}, \quad (3)$$

$$\begin{bmatrix} E_3 \\ E_4 \end{bmatrix} = \frac{E_0}{\sqrt{2}} \begin{bmatrix} H_- cos\theta - V_- sin\theta e^{i\psi} \\ i(H_- cos\xi + V_- sin\xi e^{i\psi}) \end{bmatrix}, \quad (4)$$

where $\zeta, \eta, \theta$, and $\xi$ are rotation angles of each polarizer from the horizontal axis to the counter-clockwise direction. Due to the same polarization projection of the orthogonally polarized photons onto the corresponding polarizer in each equation, the originally non-interacting photons now become coherent. Thus, the corresponding measured intensities of the non-locally paired photons are as follows:

$$I_1 = \frac{I_0}{2}(H_+ cos\zeta - V_+ sin\zeta e^{i\varphi})(H_+ cos\zeta - V_+ sin\zeta e^{-i\varphi})$$
$$= \frac{I_0}{2}[H_+ H_+ cos^2\zeta + V_+ V_+ sin^2\zeta - sin2\zeta cos\varphi],$$
$$= \frac{I_0}{2}(1 - sin2\zeta cos\varphi), \quad (5)$$

$$I_2 = \frac{I_0}{2}(H_+ cos\eta + V_+ sin\eta e^{i\varphi})(H_+ cos\eta + V_+ sin\eta e^{-i\varphi})$$
$$= \frac{I_0}{2}[H_+ H_+ cos^2\eta - V_+ V_+ sin^2\eta + sin2\eta cos\varphi],$$
$$= \frac{I_0}{2}(1 + sin2\eta cos\varphi), \quad (6)$$

$$I_3 = \frac{I_0}{2}(H_- cos\theta - V_- sin\theta e^{i\psi})(H_- cos\theta - V_- sin\theta e^{-i\psi})$$

$$= \frac{I_0}{2}[H_-H_-\cos^2\theta + V_-V_-\sin^2\theta - \sin2\theta\cos\psi],$$
$$= \frac{I_0}{2}(1 - \sin2\theta\cos\psi), \quad (7)$$
$$I_4 = \frac{I_0}{2}(H_-\cos\xi + V_-\sin\xi e^{i\psi})(H_-\cos\xi + V_-\sin\xi e^{-i\psi})$$
$$= \frac{I_0}{2}[H_-H_-\cos^2\xi + V_-V_-\sin^2\xi + \sin2\xi\cos\psi],$$
$$= \frac{I_0}{2}(1 + \sin2\xi\cos\psi). \quad (8)$$

From Eqs. (5)-(8), local measurements in each party show interference fringes, where such local fringes has already been observed even with SPDC-generated photon pairs [10]. In a single noninterfering MZI, such phenomenon is called a quantum eraser which has been observed in both SPDC [10] and coherent photon cases [32]. Thus, the function of the polarizers for the quantum eraser is successfully driven for the coherence retrieval in the noninterfereing MZIs of Fig. 1.

For the nonlocal correlation, now coincidence detection is applied to a pair of local detectors between two parties as follows via heterodyne detection to eliminate unwanted correlations, resulting in the following relations:

$$R_{14} = \frac{I_0^2}{4}(H_+\cos\zeta - V_+\sin\zeta e^{i\varphi})(H_-\cos\xi + V_-\sin\xi e^{i\psi})(H_+\cos\zeta - V_+\sin\zeta e^{-i\varphi})(H_-\cos\xi + V_-\sin\xi e^{-i\psi}). \quad (9)$$

For $\psi = \varphi = 0$, Eq. (9) results in nonlocal correlation via coincident heterodyne detection:

$$R_{14} = \frac{I_0^2}{4}(H_+H_-\cos\zeta\cos\xi - V_+V_-\sin\zeta\sin\xi - H_+V_-\cos\zeta\sin\xi - H_-\cos\xi V_+\sin\zeta)(c.c.),$$
$$= \frac{I_0^2}{4}[\cos(\zeta + \xi) - \sin(\zeta + \xi)]^2. \quad (10)$$

Likewise, $R_{23} = R_{14}$ is achieved. Thus, the nonlocal quantum correlation is successfully analyzed for Fig. 1 via polarization-frequency correlation and heterodyne detection, in which the measured inseparable intensity products are the origin of the Bell inequality violation (see Fig. 2(a)) [20,38].

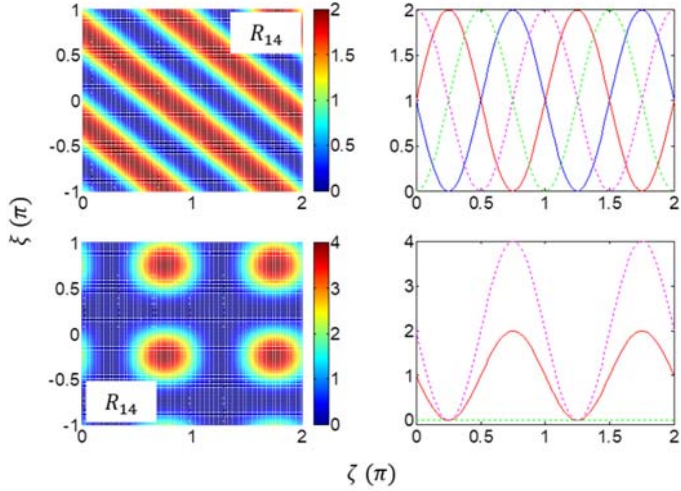

Fig. 2. Numerical calculations for Fig. 1. (Top panels) quantum feature of Eq. (10). (Bottom panels) Classical feature of Eq. (9) for $\psi = 0$ and $\varphi = \pi$. $\xi = 0$ (B); $\frac{\pi}{4}$ (G dotted); $\frac{\pi}{2}$ (R); $\frac{3\pi}{4}$ (M dotted). In the bottom panels, the red curve is also for $\xi = \frac{\pi}{4}$.

The top panels of Fig. 2 show numerical calculations of Eq. (10) or Eq. (9) for $\varphi = \psi = 0$, representing the typical nonlocal quantum feature. The upper right panel clearly shows the correlation relation between independent polarizers separated by space-like distance. The bottom panels of Fig. 2 show the classical feature of Eq. (9) for $\varphi = \pi$ and $\psi = 0$. The red curve for $\xi = \frac{\pi}{2}$ coincides with the blue curve for $\xi = 0$. Thus, coherently manipulated nonlocal quantum feature between Alice and Bob is clearly demonstrated for inseparable intensity products, where the light source is Poisson distributed coherent photons. The Bell inequality violation is with respect to the classical linear optics [20], where sinusoidal oscillation of the inseparable intensity products with a fringe visibility greater than 71% is the direct proof of the quantum feature. The degree of Bell inequality violation in the upper right panel of Fig. 2 is $S = 2\sqrt{2}$, representing a perfect nonlocal quantum feature [20,25,33,38,39].

**Conclusion**
Using coherent manipulations of polarization-frequency correlation in a noninterfering MZI via a quantum eraser, nonlocal correlation between the MZI output photons was analyzed and numerically demonstrated for Bell inequality violations. The manipulation of polarization-frequency correlation was conducted by a pair of AOMs through a PBS. The coincident heterodyne detection selectively and coincidently chose difference frequency between two local measurements, resulting in the inseparable intensity products. Unlike conventional understanding of the nonlocal correlation limited to quantum particles such as entangled photon pairs, the present classical scheme with coherent photons is counterintuitive according to the common understanding that nonlocal correlation cannot be achieved by any classical means. Unlike coincidence detection-caused measurement-basis selections in conventional Franson-type nonlocal correlation [24,25], the proposed method is for the same measurement-basis control for difference frequency via heterodyne detection, satisfying the same inseparable basis products via a quantum eraser. Regarding photon indistinguishability for the nonlocal correlation, the polarization-basis randomness in both local detectors was satisfied by the quantum eraser of the delayed-choice experiments. Here, the delayed-choice for the nonlocal quantum feature coverts orthogonally polarized photons into coherent photons. The presented nonlocal quantum feature based on a coincidence and heterodyne detection technique may be extended into a macroscopic realm of cw light, resulting in macroscopically excited nonlocal correlation, if the inseparable intensity products can be achieved macroscopically. The measurement choice by heterodyne detection played a critical role for the coherently excited nonlocal quantum feature.

**Acknowledgments**
This work was supported by ICT R&D program of MSIT/IITP (2021-0-01810), development of elemental technologies for ultrasecure quantum internet and the GIST Research Project in 2022.